\documentclass[twocolumn,showpacs]{revtex4}

\usepackage{amsfonts}
\usepackage{amssymb}
\usepackage{amsbsy}
 \usepackage{bm}

\begin{document}

\title{Understanding multilayers from a 
geometrical viewpoint}

\author{T. Yonte, J. J. Monz\'on, L. L. S\'anchez-Soto}
\affiliation{Departamento de \'Optica, 
Facultad de Ciencias F\'{\i}sicas, 
Universidad Complutense, 
28040 Madrid,  Spain}

\author{J. F. Cari\~{n}ena}
\affiliation{Departamento de F\'{\i}sica Te\'orica, 
Facultad de Ciencias, Universidad de Zaragoza, 
50009 Zaragoza,  Spain}

\author{C. L\'opez-Lacasta}
\affiliation{Departamento de Matem\'{a}tica 
Aplicada, CPS, Universidad  de Zaragoza,
50015 Zaragoza, Spain.}

\date{\today}

\begin{abstract}
We reelaborate on the basic properties of 
lossless multilayers. We show that the transfer
matrices for these multilayers have essentially
the same algebraic properties as the Lorentz
group SO(2,1) in a $(2+1)$-dimensional
spacetime, as well as the group SL(2,$\mathbb{R}$)
underlying the structure of the $ABCD$ law
in geometrical optics. By resorting to the Iwasawa
decomposition, we represent the action of any 
multilayer as the product of three matrices 
of simple interpretation. This group-theoretical
structure allows us to introduce bilinear 
transformations in the complex plane. The concept
of multilayer transfer function naturally emerges
and its corresponding properties in the unit disc
are studied. We show that the Iwasawa 
decomposition reflects at this geometrical level
in three simple actions that can be considered
the basic pieces for a deeper undestanding of
the multilayer behavior. We use the method to 
analyze in detail a simple practical example.
\end{abstract}

\pacs{120.5700 Reflection, 120.7000 Transmission, 
230.4170 Multilayers, 000.3860 Mathematical 
methods in physics}

\maketitle

\section{Introduction}

Layered media are important for many applications 
in modern optics. To fully capitalize these media 
one must have a clear picture of all the mechanisms 
involved in the propagation of the optical waves 
in them. In spite of this fact, multilayer optics 
is usually approached from a practical view, 
in relation to optical filters and the like. In this 
spirit, the topics covered in most of the textbooks 
on the subject use a mixture of design, manufacture, 
and applications; treating only the basic physics needed 
to carry out  practical computations~\cite{MA86}.

However, for a variety of reasons layered media have 
physical relevance on their own~\cite{YE88,LE87}. 
The mathematical basis for understanding their behavior is 
the fact that the matrix representing any lossless multilayer 
belongs to the group SU(1,1). It is known that this group is 
locally isomorphic to the $(2+1)$-dimensional  Lorentz group 
SO(2,1)~\cite{MO99c,MO99a}. This leads to a natural
and complete identification between reflection and 
transmission coefficients of the multilayer and the 
parameters of the corresponding Lorentz transformation. 
It is, precisely, the abstract composition law of SU(1,1)  
the ultimate responsible for the curious composition 
law of these reflection and transmission coefficients. 
Moreover, this fact allows one to perform experimental 
tests of special relativity with simple optical measurements  
involving multilayers~\cite{MO99b,MO99d,MO01a}.

In this respect, another remarkable fact to be considered 
is that SU(1,1) is also isomorphic to SL(2,$\mathbb{R}$),
which is the natural group underlying the mathematical
structure of the celebrated $ABCD$ law in first-order 
optics. In this paper, we also exploit this correspondence 
to explore intriguing connections of layered media
with geometrical optics.

These purely algebraic results seem to call for a geometrical
interpretation. It is difficult to overestimate the role played
by geometrical ideas in all the branches of physics, 
particularly in special relativity. The hyperbolic geometry 
associated with the group SU(1,1) [or, equivalently, SO(2,1)] is 
an established doctrine~\cite{CO68}. In light of these
considerations, it is easy to convince oneself that this 
geometrical approach might provide deeper insights into 
the action of a multilayer in a wider unifying framework 
that can put forward fruitful analogies with other 
physical phenomena.

In consequence, it is natural to view the action of 
a SU(1,1) multilayer matrix as a bilinear (or M\"{o}bius) 
transformation on the unit disc, obtained by stereographic 
projection of the unit hyperboloid of SO(2,1). This kind of bilinear 
representations have been discussed in detail for the Poincar\'e 
sphere in  polarization optics~\cite{AZ87,HA96}, for Gaussian 
beam propagation~\cite{KO65}, and are also useful in laser 
mode-locking and optical pulse transmission~\cite{NA98}.

In addition, the isomorphism with SL(2,$\mathbb{R}$)
allows us to translate the geometrical structure defined in
the unit disc to the complex plane, recovering in this way
an alternative model of the hyperbolic geometry that
is useful in shedding light into the  behavior of the multilayer. 

In spite of these achievements, the geometrical action of an
arbitrary lossless multilayer could still become cumbersome 
to interpret in physical terms. In fact, in practice it is usual 
to work directly with  the numerical values of a matrix 
obtained from the experiment, which cannot be directly 
related to the inner multilayer structure. To remedy this 
situation, we have resorted recently~\cite{MO01b}
to the Iwasawa decomposition, which provides a remarkable
factorization of the matrix representing any multilayer 
(no matter how complicated it could be)  as  the product  
of three matrices of simple interpretation.

At the geometric level, such a decomposition translates
directly into the classification of three basic actions, which 
are studied in this paper, that are the basic bricks from which 
any multilayer action is built.

The contents of this paper are organized as follows.
In Section 2 we present in detail some relevant
algebraic properties of lossless multilayers. These 
properties have a direct translation into hyperbolic 
geometry, both in the unit disc and in the complex plane, 
which is explored in Section 3. Finally, Section 4 devote 
to develop a simple yet relevant example that illustrates 
the power of this approach and to expose our concluding 
remarks.

\section{Some properties of  lossless multilayer matrices}

\subsection{Transfer matrix for a lossless multilayer}

We first briefly summarize the essential ingredients 
of multilayer optics we shall need for our purposes~\cite{AZ87}.
The  configuration is a stratified structure, illustrated 
in Fig.~1, that consists of a stack of  $1, \ldots, j, \ldots, m$,  
plane--parallel lossless layers sandwiched 
between two semi--infinite ambient ($a$) and substrate 
($s$) media, which we shall assume to be identical, since
this is the common experimental case. Hereafter
all the media are supposed to be lossless,  
linear, homogeneous, and isotropic. 

We consider an incident monochromatic linearly 
polarized plane wave from the ambient, which makes 
an angle  $\theta_0$ with  the normal to the first 
interface and  has amplitude  $E_{a}^{(+)}$. 
The electric field is either in the plane of  incidence 
($p$ polarization) or perpendicular to the plane of 
incidence ($s$ polarization). We consider as well 
another plane wave of the same frequency and  
polarization, and with amplitude $E_{s}^{(-)}$, 
incident from the substrate at the same  angle 
$\theta _{0}$.~\cite{Snell}

As a result of multiple reflections in all the interfaces,
we have a backward-traveling plane wave in the 
ambient, denoted $E_{a}^{(-)}$, and a  
forward-traveling plane wave in the substrate, 
denoted $E_{s}^{(+)}$. If we consider  the field  
amplitudes as a vector of the form
\begin{equation}
\label{Evec}
\mathbf{E} = 
\left ( \begin{array}{c}
E^{(+)} \\ 
E^{(-)} 
\end{array}
\right )\ ,
\end{equation}
which applies to both ambient and substrate
media, then the amplitudes at each side 
of the multilayer are related by a $2 \times 2$ 
complex matrix $\mathsf{M}_{as}$, we shall 
call the multilayer transfer matrix~\cite{OH00}, 
in the form 
\begin{equation}
\label{M1}
\mathbf{E}_a =  
\mathsf{M}_{as} \,
\mathbf{E}_s\ .
\end{equation}
The matrix  $\mathsf{M}_{as}$ can be shown to 
be~\cite{MO99c} 
\begin{equation}
\label{Mlossless}
\mathsf{M}_{as} =
\left [
\begin{array}{cc}
1/T_{as} & R _{as}^\ast/T_{as}^\ast \\ 
R_{as}/T_{as} & 1/T_{as}^\ast
\end{array}
\right ]  
\equiv
\left [
\begin{array}{cc}
\alpha & \beta \\ 
 \beta^\ast & \alpha^\ast
\end{array}
\right ]  ,
\end{equation}
where the complex numbers 
\begin{equation}
R_{as}  =  | R_{as} | \exp (i \rho), 
\qquad  
T_{as}  =  | T_{as} | \exp (i \tau) ,
\end{equation} 
are, respectively,  the overall reflection and transmission 
coefficients for a wave incident from the ambient.
Note that  
\begin{equation}
\label{hyperb}
\det \mathsf{M}_{as}=|\alpha|^2 -|\beta|^2 =
\frac{1-|R_{as}|^2}{|T_{as}|^2} = +1 . 
\end{equation}
Therefore, the condition $\det \mathsf{M}_{as}= +1$ 
is equivalent to  $|R_{as}|^2  + |T_{as}|^2 = 1$, 
and then  the set of lossless multilayer matrices
reduces to the group SU(1,1), whose elements depend
on three independent real parameters.

When the ambient and substrate media are different, 
this result also holds after a convenient  renormalization 
of  the field amplitudes~\cite{MO99c}. The identity matrix 
corresponds to $T_{as} =1$ and $R_{as} = 0$, so it 
represents an antireflection system.  The matrix that
describe the overall system obtained by putting two multilayers
together is the product of the matrices representing each 
one of them, taken in the appropriate order.  So, two 
multilayers, which are inverse, when composed give 
an antireflection system. 

\subsection{A basic factorization for multilayers: 
the Iwasawa decomposition}
 
Many types of matrix factorizations have been
considered in the literature~\cite{AR83, AB94,SH95}, 
all of them decomposing the matrix as a unique 
product  of other matrices of simpler interpretation. 
Particularly, given the essential role  played by  the 
Iwasawa decomposition, both in fundamental 
studies and in applications  to several fields (especially 
in  optics), one is tempted to investigate also its role in 
multilayer  optics.  

Without embarking us in mathematical subtleties,  the  
Iwasawa decomposition is established  as follows~\cite{BA77,HE78}: 
any element $g$ of a (noncompact semi-simple) Lie group 
can be written  as an ordered product of three elements, 
taken one each from a maximal compact subgroup 
$K$, a maximal Abelian subgroup $A$, and  a maximal 
nilpotent subgroup $N$. Furthermore, such a  decomposition 
is global (in the sense that it applies to every group element)
and essentially unique (in the sense that the elements of the 
factorization are uniquely determined in terms of $g$). 

For the problem at hand of  a lossless multilayer matrix 
$\mathsf{M}_{as}  \in$  SU(1,1),  the decomposition reads 
as \cite{MO01b}
\begin{equation}
\label{Iwa1}
\mathsf{M}_{as} = \mathsf{K}(\phi)\,  \mathsf{A}(\xi)\, 
\mathsf{N}(\nu)\ ,
\end{equation}
where
\begin{eqnarray}
\label{Iwasa1}
\mathsf{K}(\phi) & = & 
\left [
\begin{array}{cc}
\exp (i\phi/2) & 0 \\ 
0 & \exp (-i\phi/2)
\end{array}
\right ]\  , 
\nonumber \\
\mathsf{A}(\xi) & = & 
\left [
\begin{array}{cc}
\cosh (\xi/2) & i\, \sinh(\xi/2) \\ 
-i\, \sinh(\xi/2) & \cosh (\xi/2)
\end{array}
\right ]\ , \\
\mathsf{N}(\nu) & = & 
 \left [
\begin{array}{cc}
1 - i\, \nu/2& \nu/2 \\ 
\nu/2 & 1+ i\, \nu/2
\end{array}
\right ]\  .
\nonumber
\end{eqnarray}
The parameters $\phi, \xi$,
and $\nu$ are given in terms of the
elements of the multilayer matrix by
\begin{eqnarray}
\label{param}
\phi/2 & = & \arg (\alpha + i \beta)\ ,  \nonumber  \\
\xi/2  &  =  & \ln  (1/|\alpha +i \beta | )\ , \\
\nu/2 & = & \mathrm{Re} (\alpha  \beta^\ast)/
|\alpha + i \beta |^2\ ,  \nonumber 
\end{eqnarray}
where the ranges of the parameters are  $\xi,  \nu \in 
\mathbb{R}$ and $-2\pi \le \phi \le 2\pi$. Therefore, 
given \textit{a priori} limits on $\alpha$ and 
$\beta$ (i.e., on $T_{as}$ and $R_{as}$), one could
easily establish the corresponding limits on the 
parameters $\phi$, $\xi$ and $\nu$, and vice versa.

All the matrices in Eq.~(\ref {Iwa1}) are of SU(1,1)  
and, therefore,  leave invariant the expression 
$|E^{(+)}|^2 - |E^{(-)}|^2$ at each side of the multilayer; i.e.,
\begin{equation} 
\label{flujo}
|E^{(+)}_a|^2 - |E^{(-)}_a|^2 = 
|E^{(+)}_s|^2 - |E^{(-)}_s|^2\ ,
\end{equation}
which is nothing but the energy-flux conservation, as 
could be expected from physical considerations. 
In addition, the matrix $\mathsf{K}(\phi)$ preserves 
the product $E^{(+)}E^{(-)}$,  the matrix  $\mathsf{A}(\xi) $  
preserves the quadratic form ${E^{(+)}}^2 + {E^{(-)}}^2$, 
and  the matrix $\mathsf{N}(\nu)$ preserves the sum 
$E^{(+)}+ i\,E^{(-)}$. 

\subsection{Connections with geometrical optics}

It this subsection we wish to show a remarkable analogy 
with geometrical optics that could provide new insights 
into the interpretation of the action of any lossless
multilayer. To this end we note that by applying the 
unitary matrix 
\begin{equation}
\label{Unit}
{\mathcal{U}} =
\frac{1}{\sqrt{2}}
\left [
\begin{array}{cc}
1 & i \\ 
i & 1
\end{array}
\right ]  
\end{equation}
to both sides of Eq.~(\ref{M1}), we can recast it as
\begin{equation}
\bm{{\mathcal{E}}}_a =  
{\mathcal{M}}_{as} \,
\bm{{\mathcal{E}}}_s\ ,
\end{equation}
where the new field vectors are defined as
\begin{equation}
\bm{{\mathcal{E}}} =  
\left ( \begin{array}{c}
{\mathcal{E}}^{(+)} \\ 
{\mathcal{E}}^{(-)} 
\end{array}
\right ) =
{\mathcal{U}}\, \mathbf{E} = 
\frac{1}{\sqrt{2}}
\left ( \begin{array}{c}
E^{(+)}  + i E^{(-)}  \\ 
E^{(-)} + i E^{(+)}  
\end{array}
\right ) ,
\end{equation}
and the conjugate multilayer matrix is
\begin{equation}
{\mathcal{M}}_{as} =  
{\mathcal{U}}\, {\mathsf{M}}_{as}\,{\mathcal{U}}^{-1} =
\left [ 
\begin{array}{cc}
a & b \\ 
c & d
\end{array}
\right ]\ , 
\end{equation}
where ${\mathcal{M}}_{as}$ is a matrix with 
$\det{\mathcal{M}}_{as} = +1$ and whose elements 
are  real numbers given by
\begin{eqnarray}
a =  {\rm Re} (\alpha) + {\rm Im} (\beta)\,,
\qquad 
b = {\rm Im}  (\alpha) + {\rm Re} (\beta)\,, 
\nonumber \\
& & \\
c = -  {\rm Im}  (\alpha) + {\rm Re} (\beta) \, ,
\qquad
d = {\rm Re} (\alpha) -   {\rm Im} (\beta) \, .
\nonumber 
\end{eqnarray}
In other words, the matrices ${\mathcal{M}}_{as}$ belong 
to the group SL(2,$\mathbb{R}$), which plays an 
essential role in a variety of branches in optics, ranging 
from the celebrated $ABCD$ law of geometrical 
optics~\cite{geom1,geom2,geom3,geom4},  
to squeezed states in quantum optics~\cite{ssqo1,ssqo2}, 
including beam propagation problems~\cite{bpp}. 
The transformation by ${\mathcal{U}}$ establishes 
a one-to-one map between the group  SL(2,$\mathbb{R}$) 
of  matrices  ${\mathcal{M}}_{as}$ and the group SU(1,1) 
of matrices ${\mathsf{M}}_{as}$, which allows for a 
direct  translation of the properties from one to the other. 
Such a correspondence is an isomorphism of groups, because 
\begin{equation}
{\mathcal{M}}^{(2)}_{as}\,{\mathcal{M}}^{(1)}_{as}=
{\mathcal{U}}\, {\mathsf{M}}^{(2)}_{as}\,
{\mathcal{U}}^{-1}\,{\mathcal{U}}\, 
{\mathsf{M}}^{(1)}_{as}\,{\mathcal{U}}^{-1}=
{\mathcal{U}}\,{\mathsf{M}}^{(2)}_{as}\,
{\mathsf{M}}^{(1)}_{as}\,
{\mathcal{U}}^{-1}\ .
\end{equation}

By conjugating with ${\mathcal{U}}$ the Iwasawa 
decomposition (\ref{Iwa1}), we get  the corresponding 
one for SL(2,$\mathbb{R}$), which has been 
previously worked out~\cite{Iwasl2r}:
\begin{equation}
\label{Iwa2}
{\mathcal{M}}_{as} = 
{\mathcal{K}}(\phi)\,  
{\mathcal{A}}(\xi)\, 
{\mathcal{N}}(\nu)\ ,
\end{equation}
where
\begin{eqnarray}
{\mathcal{K}}(\phi)  & = & 
\left [
\begin{array}{cc}
\cos(\phi/2) &  \sin(\phi/2) \\ 
- \sin(\phi/2)  & \cos(\phi/2) 
\end{array}
\right ] \ , 
\nonumber \\
{\mathcal{A}}(\xi) & = & 
\left [
\begin{array}{cc}
\exp(\xi/2) & 0 \\ 
0 & \exp (-\xi/2)
\end{array}
\right ]\ , \\
{\mathcal{N}}(\nu) & = & 
 \left [
\begin{array}{cc}
1 & 0 \\ 
 \nu & 1
\end{array}
\right ]\ .
\nonumber
\end{eqnarray}

Now, we can interpret the physical action of the 
matrices appearing in both factorizations in SU(1,1) and 
SL(2,$\mathbb{R}$), respectively. In this way, 
${\mathsf{K}}(\phi)$  represents the free propagation 
of the fields  $\mathbf{E}$ in the ambient medium through
an optical phase thickness of $\phi/2$. Obviously, 
this reduces to a mere shift of the origin of phases.  
Alternatively,  one can consider ${\mathcal{K}}(\phi)$ as 
an $ABCD$ matrix in geometrical optics that applies
to position $\mathbf{x}$ and momentum $\mathbf{p}$
(direction) coordinates of a ray in a transverse 
plane~\cite{CN}. These are the natural phase-space 
variables of ray optics and then ${\mathcal{K}}(\phi)$  
would represent a rotation in these variables~\cite{bpp}.
In the multilayer picture, ${\mathcal{E}}^{(+)}$ 
can be seen as the variable  $\mathbf{x}$
and ${\mathcal{E}}^{(-)}$ can be seen as
the corresponding $\mathbf{p}$.

In Eq.~(\ref{Iwasa1}), the second matrix 
${\mathsf{A}}(\xi)$  represents a symmetric 
multilayer (i.e., the reflection coefficient is the 
same whether light is incident on one side or on 
the opposite side of the multilayer,  and so 
$\tau_{\mathsf{A}} - \rho_{\mathsf{A}} = \pm \pi/2$) 
with transmission and reflection phase shifts of 
$\tau_{\mathsf{A}} = 0$ and $\rho_{\mathsf{A}} =  
\pm \pi/2$, and a transmission coefficient 
$T_{\mathsf{A}} = \mathrm{sech}(\xi/2)$.  
There are many ways to get this performance, 
perhaps the simplest one is a Fabry-Perot 
system composed by two identical plates separated
by a transparent spacer. By adjusting the 
refractive indices and the thicknesses of the 
media one can always get the desired values.
Viewed  in  SL(2,$\mathbb{R}$),  ${\mathcal{A}}(\xi)$ 
represents a magnifier that scales $\mathbf{x}$ 
up to the factor $m = \exp(\xi/2)$ and $\mathbf{p}$ 
down by the same factor~\cite{bpp}. 

The third matrix, ${\mathsf{N}}(\nu)$, represents a 
system having $T_{\mathsf{N}} = 
\cos(\tau_{\mathsf{N}}) \exp(i \tau_{\mathsf{N}})$ 
and $R_{\mathsf{N}} =  \sin(\tau_{\mathsf{N}}) 
\exp(i \tau_{\mathsf{N}})$,  with  $ \tan
(\tau_{\mathsf{N}}) =  \nu/2$. The simplest way 
to accomplish this task is by an asymmetrical 
two-layer system. Using the analogy with the $ABCD$ 
matrix in geometrical optics, ${\mathcal{N}}(\nu)$ 
represents the action of a lens of power 
$\nu$~\cite{bpp}.  

Finally, in complete equivalence with the invariants
found for the Iwasawa decomposition of SU(1,1), the
matrix ${\mathcal{K}}$ preserves the sum 
${{\mathcal{E}}^{(+)}}^2 + {{\mathcal{E}}^{(-)}}^2$, 
${\mathcal{A}}$ preserves the product 
${\mathcal{E}}^{(+)} {\mathcal{E}}^{(-)}$, and
${\mathcal{N}}$ preserves ${\mathcal{E}}^{(+)}$.
In addition, the energy-flux conservation Eq.~(\ref{flujo}) 
can be recast now as the invariance of 
$\mathrm{Im} [ {\mathcal{E}}^{(+)} 
{{\mathcal{E}}^{(-)}}^\ast  ]$.

\subsection{Connections with special relativity}

Apart from the geometrical-optics perspective 
developed in the previous subsection, multilayer 
action can be viewed in a relativisticlike framework,
that has proved to be very appropriate to understand
some peculiarities of multilayer behavior. To this
end, let us first recall some well-known facts about the  
Lorentz transformations in (2+1)-dimensions.

Introducing a three-dimensional real vector space 
of vectors with components $(x^0, x^1, x^2)$, 
where $x^0 = ct$, a Lorentz transformation 
$\Lambda$ is a linear transformation between 
two coordinate frames
\begin{equation}
{x^\prime}^\mu = \Lambda^\mu\,_\nu \ x^\nu 
\end{equation}
(the Greek indices run from 0 to 2), such that the  
pseudo-Euclidean bilinear form 
\begin{equation}
\langle x | y \rangle =
x^0 y^0 - x^1 y^1 - x^2 y^2 
\label{Lorentzpr}
\end{equation}
remains invariant; i.e., $\langle x^\prime | y^\prime 
\rangle =  \langle x | y \rangle$, which immediately 
implies
\begin{equation}
\det \Lambda = \pm 1 .
\end{equation}
Thus,  the set of Lorentz transformations can be 
classified in two classes: proper transformations, with 
$\det \Lambda = + 1$, and improper  ones, with 
$\det \Lambda = - 1$. The proper Lorentz transformations 
form a subgroup, but  the improper ones do not. 

Furthermore, it is easy to check that $|\Lambda^0\,_0|\geq 1$.  
Therefore, the transformations of the  Lorentz group can also 
be classified according to the sign of $\Lambda^0\,_0$:  the 
orthochronous Lorentz  transformations, with $\Lambda_0^0 
\ge 1$, form a subgroup, but the antichronous ones, with 
$\Lambda^0\,_0 \le -1$, do not. 

We are interested in dealing with the three-parameter 
restricted  Lorentz group SO(2,1); i.e., the group of 
the Lorentz transformations with determinant $+ 1$ and 
that do not reverse the direction of time. In fact, we
wish to discuss now a very close correspondence 
between the group  SU(1,1) introduced above and 
the restricted Lorentz  group  SO(2,1).  We shall 
show this important correspondence explicitly  
in the following form that we recall for clarity~\cite{BA47}:  
with each point of our three-dimensional vector 
space with coordinates $x^\mu$ we associate  
the Hermitian matrix
\begin{equation}
\mathsf{X} = x^\mu \sigma_\mu =
\left [ 
\begin{array}{cc}
x^0 & x^1- i\,x^2 \\ 
x^1 + i\, x^2 & x^0
\end{array}
\right ] ,  \label{ftalrel}
\end{equation}
where $\sigma_0 = I$ is the identity, and $\sigma_1$
and  $\sigma_2$ are the corresponding Pauli matrices.  
Note, that $\det \mathsf{X}= \langle x | x \rangle= 
(x^0)^2 - (x^1)^2 -(x^2)^2$. 

Now, if  $\mathsf{M} \in $ SU(1,1), then the matrix  
\begin{equation}
\mathsf{X}^\prime = \mathsf{M} \
\mathsf{X} \
\mathsf{M}^\dagger ,
\label{suunoact}
\end{equation}
where the symbol $\dagger$ denotes the Hermitian 
conjugate,  induces a Lorentz transformation on the 
coefficients $x^\mu$.

It is clear from Eq.~(\ref{suunoact}) that the matrices 
$\mathsf{M}$ and  $- \mathsf{M}$ generate the same 
$\Lambda$,  so this homomorphism is two-to-one. 
This equation can be easily solved to obtain $\mathsf{M}$ 
(uniquely defined  up to the sign) from a given $\Lambda$.
 In fact, it is easy to find it explicitly as~\cite{BA77}
\begin{equation}
\label{LM}
\Lambda^\mu\,_\nu (\mathsf{M}) = 
\frac{1}{2} {\rm Tr} \left ( \sigma^\mu \mathsf{M} 
\sigma_\nu \mathsf{M}^\dagger \right )\ .
\end{equation}  

While $\mathsf{M}$ acts on two-dimensional
complex vectors like those in Eq.~(1), the induced 
Lorentz transformation $\Lambda (\mathsf{M})$
acts on three-dimensional real vectors of the form
[which are the \textit{space-time} counterparts of
Eq.~(1)]:
\begin{equation}
\left( 
\begin{array}{c}
x^0  \\ 
x^1 \\
x^2 
\end{array}
\right) 
\rightarrow
\left (
\begin{array}{c}
e^0 \\
e^1 \\
e^2 \\ 
\end{array}
\right) 
= 
\left (
\begin{array}{c}
(|E^{(+)}|^2 + |E^{(-)}|^2)/2 \\
{\rm Re} [{E^{(+)}}^\ast E^{(-)} ]\\
{\rm Im} [{E^{(+)}}^\ast E^{(-)} ] 
\end{array}
\right) . 
\end{equation}
The temporal coordinate is the semi-sum of the 
fluxes at each side  of the multilayer. The interval 
remains invariant 
\begin{equation}
\label{intervalo}
(e^0)^2 - (e^1)^2 - (e^2)^2  = K^2.
\end{equation}
This number $K$ (which is the `radius') is the 
semi-difference of the fluxes at each side of 
the multilayer;  i.e., $(|E^{(+)}|^2 - |E^{(-)}|^2 )/4$ 
for both ambient or substrate and, therefore,
it can take any real value. Without loss of generality 
we can renormalize these variables so as to take the
value of $K$ equal to 1 and then we are working 
on the unit two-sheeted hyperboloid of SO(2,1)~\cite{MI88}. 

In summary, given the multilayer matrix 
${\mathsf{M}}_{as}$ in Eq.~(\ref{Mlossless}), 
the corresponding Lorentz transformation
in SO(2,1) is~\cite{MO99c}
\begin{equation}
\label{Lorentz}
\Lambda (\mathsf{M}_{as}) =
\left [ 
\begin{array}{ccc}
|\alpha|^{2} + |\beta|^{2} & 2 {\rm  Re} (\alpha\beta^\ast)
& 2 {\rm Im}(\alpha\beta^ \ast) \\
2  {\rm Re} (\alpha \beta ) & {\rm Re} (\alpha^2 + \beta^2) 
& {\rm Im} (\alpha^2 - \beta^2 ) \\
- 2  {\rm Im} (\alpha \beta )  &
-{\rm Im} ( \alpha^2 + \beta^2 ) & 
{\rm Re} ( \alpha^2 - \beta^2 ) 
\end{array}
\right ] . 
\end{equation}
Moreover, using Eq.~(\ref{LM}) it is direct to obtain
the explicit expressions in SO(2,1) for each matrix appearing 
in the Iwasawa decomposition (\ref{Iwa1}). Indeed one finds 
\begin{eqnarray}
\Lambda_{\mathsf{K}} (\phi ) 
& = &
\left [ 
\begin{array}{ccc}
1 &  0 & 0 \\ 
0 &   \cos \phi &   \sin \phi \\
0 & -\sin \phi &  \cos \phi
\end{array}
\right ] , \nonumber \\
\Lambda_{\mathsf{A}} (\xi) 
& = & 
\left [ 
\begin{array}{ccc}
\cosh \xi &  0 & -\sinh \xi \\ 
0 &   1 & 0 \\
-\sinh \xi & 0 &  \cosh \xi
\end{array}
\right ] ,  \\
\Lambda_{\mathsf{N}} (\nu) 
& = &
\left [ 
\begin{array}{ccc}
1 + (\nu^2/2) &  \nu & -\nu^2/2 \\ 
\nu &   1 & -\nu \\
\nu^2/2 & \nu &  1 - (\nu^2/2)
\end{array}
\right ] .  \nonumber 
\end{eqnarray}
The action of these matrices in SO(2,1) is
clear: $\Lambda_{\mathsf{K}} (\phi )$ is 
a space rotation of angle $\phi$ in the
$e^1 - e^2$ plane, $\Lambda_{\mathsf{A}} (\xi)$
is a boost in the direction of the axis $e^2$ with  
velocity $v/c = \tanh \xi$; and, finally, 
$\Lambda_{\mathsf{N}} (\nu)$ represents a 
space rotation of angle $\tau_{\mathsf{N}}$ 
[such that $\tan (\tau_{\mathsf{N}}) = \nu/2$] followed 
by a boost of angle $\tau_{\mathsf{N}}$ and velocity
$v/c = \tanh (\nu/2)$, both in the $e^1 - e^2$ plane.
In the next Section we shall explore in more detail
the properties of these three matrices.

\section{Geometrical interpretation of the multilayer action}

\subsection{Multilayer transfer function and hyperbolic 
geometry  in the unit disc}

In many instances (e.g., in polarization optics~\cite{AZ87}) 
we are interested in the transformation properties of field 
quotients rather than the fields themselves. Therefore, 
it seems natural to consider the complex numbers
\begin{eqnarray}
\label{defz}
z_s   =  \frac {E_s^{(-)}}{E_s^{(+)}}\ ,
\qquad
z_a  =   \frac {E_a^{(-)}}{E_a^{(+)}}\ .
\end{eqnarray}
The action of the multilayer given in Eq.~(\ref{Mlossless}) 
can be then seen as a function  $z_a = f(z_s)$ that can 
be appropriately called the multilayer transfer 
function~\cite{OH00}.

From a geometrical viewpoint, this function defines a 
transformation of  the complex plane ${\mathbb{C}}$,
mapping the point $z_s$ into the point $z_a$, according to
\begin{equation}
\label{accion}
z_a = \Phi [\mathsf{M}_{as} , z _{s}] = 
\frac{\beta^\ast +\alpha^\ast z_s} 
{\alpha + \beta z_s} \ ,
\end{equation}
and the point of the infinity is given by
\begin{equation}
\Phi [\mathsf{M}_{as}, -\alpha/\beta ] = \infty \ ,
\qquad 
\Phi [ \mathsf{M}_{as},\infty ] = \alpha^\ast/\beta \ .
\end{equation}
This bilinear transformation defines an action of the 
group SU(1,1) of multilayer transfer matrices on 
the complex plane $\mathbb{C}$. The complex 
plane appears then foliated  in three regions that 
remain invariant under the action of the group: the  
unit disc, its boundary and the external region. In fact, 
\begin{equation}
|z_a|^2 = \frac{|\alpha|^2 |z_s|^2
+|\beta|^2 + 2 \mathrm{Re} (\beta \alpha^\ast z_s)}
{|\alpha|^2 + |\beta|^2 |z_s|^2+
2 \mathrm{Re} (\beta \alpha^\ast z_s)}\ ,
\end{equation}
and, in consequence, the difference between the 
numerator and the denominator is  
\begin{equation}
(|\alpha|^2 - |\beta|^2) (|z_s|^2-1) = 
|z_s|^2-1\ ,
\end{equation}
which shows our assertion relative to the invariance
of those regions.

Alternatively, the unit disc can be seen as obtained from 
the  upper sheet of the unit two-sheet hyperboloid defined by 
Eq.~(\ref{intervalo}) with $K=1$ by means of stereographic 
projection using the south pole $(-1, 0, 0)$ as projection centre.
In fact, a simple calculation shows that, in such a case,
the projection of the point $(e^0, e^1, e^2)$ becomes in
the complex plane
\begin{equation}
z = \frac{e^1+ i e^2}{1 + e^0}= \frac {E^{(-)}}{E^{(+)}} \ ,
\end{equation}
in full agreement with Eq.~(\ref{defz}).

The boundary of the unit disc corresponds to the 
projection of  the infinity point and then, since  
$|z_a| = |z_s|=1$, it can be identified with the action 
of perfect mirrors (i.e., $T_{as} =0$).

The Iwasawa decomposition has an immediate 
translation in this geometrical framework, and one is led 
to treat separately the action of each one of the 
matrices appearing in this decomposition. To
this end, it is worth noting that the groups we are 
considering  appear always as groups of transformations 
of some space. The concept of orbit is especially 
appropriate for obtaining an intuitive meaning 
of the corresponding action. We recall that, given 
a point $P$,  its orbit is the set of  points $P^\prime$
obtained from $P$ by the action of  all the elements of 
the group. In Fig.~2 we have plotted a typical example
of the orbits for each one of the subgroups of  matrices 
$\Lambda_{\mathsf{K}}(\phi)$, $\Lambda_{\mathsf{A}}(\xi)$, 
and $\Lambda_{\mathsf{N}} (\nu)$. For $\Lambda_{\mathsf{K}}
(\phi)$ the orbits are the intersection of the hyperboloid with 
planes $e^0= \mathrm{constant}$, for $\Lambda_{\mathsf{A}}
(\xi)$   with planes $e^1= \mathrm{constant}$, and
for $\Lambda_{\mathsf{N}}(\nu)$  with planes 
$e^0 - e^2 = \mathrm{constant}$.

Through stereographic projection, as indicated in Fig.~2, 
we are working in the unit disc and the corresponding orbits  
for the SU(1,1) matrices $\mathsf{K}$,  $\mathsf{A}$, and 
$\mathsf{N}$ are
\begin{eqnarray}
z^\prime & = &  \Phi [\mathsf{K} (\phi), z ]
= z\,  \exp (-i\phi)\ , \nonumber \\
z^\prime & = & \Phi [\mathsf{A}(\xi), z ]  
=  \frac{z - i\, \tanh(\xi/2)}{1 + i\, z  \tanh(\xi/2)} \  , \\
z^\prime & =  & \Phi [ \mathsf{N} (\nu), z ]=  
\frac{z +(1+iz)\nu/2}{1 + (z - i)\nu/2}  . \nonumber
\end{eqnarray}
As plotted in Fig.~3.a, for matrices $\mathsf{K}$ the 
orbits are circumferences centered at the origin. 
Since for these matrices $R_{as}=0$, the action
of any antireflection system can be always pictured 
as one of these circumferences. For the matrices 
$\mathsf{A}$, they are arcs of circumference  
centered in the real axis and going from the point $ +i$ 
to the point $-i$ through $z$. Finally, for the matrices
$\mathsf{N}$ the orbits are circumferences centered 
in the imaginary axis and passing all of them through 
the points $i$, $z$, and $-z^\ast$.

The importance of the Iwasawa decomposition reflects
also at the geometrical level: no matter how complicated 
a multilayer is, its action can always be viewed in terms
of these three basic actions with a clear geometric
meaning. Its explicit application to a real case will be
demonstrated in the next Section.

\subsection{Hyperbolic geometry in the upper semi-plane}

The unitary transformation  (\ref{Unit}) has played an
important role as intertwining between multilayer and
geometrical optics. One can expect that the structure
defined in the unit disc for the former could be translated
accordingly for the latter. To this end, note that if the 
point $w\in \mathbb{C}$ is defined  in terms of $z$ by 
\begin{equation}
w = \Phi [{\mathcal{U}}, z ]  =  \frac{z+i}{1 + i\,z} \ ,
\end{equation}
it is easy to check that  the interior of the unit disc is 
mapped onto the upper semi-plane of the complex plane 
$w$, the boundary maps onto the real axis, while the 
exterior of the unit disc becomes the lower semi-plane.

The relationship between SU(1,1) and SL(2,$\mathbb{R}$) 
outlined in Section 2 allows us to transport the action 
of SU(1,1) onto $\mathbb{C}$  to give an action $\Psi$
of SL(2,$\mathbb{R}$) onto $\mathbb{C}$, finding in this way 
an  alternative model of the hyperbolic geometry and  
for determining the orbits of the matrices of 
SL(2,$\mathbb{R}$), that are the natural arena of 
geometrical optics. 

The corresponding orbits are now
\begin{eqnarray}
w^\prime & = & \Psi [{\mathcal{K}}(\phi), w ] =  
\frac{w - \tan(\phi/2)}{1 + w \tan(\phi/2)}\ , \nonumber \\
w^\prime & = &  \Psi [{\mathcal{A}}(\xi) , w ]
=  w \exp (-\xi)\ , \\
w^\prime & = &  \Psi [{\mathcal{N}}(\nu), w ]
= w+ \nu . \nonumber 
\end{eqnarray}
For matrices ${\mathcal{K}}$ the orbits are 
circumferences centered in the imaginary axis 
passing through $w$ and $-1/w$. For the matrices  
${\mathcal{A}}$, they are straight lines in the 
upper semi-plane passing through the origin and 
the point $w$. Finally, for the matrices  ${\mathcal{N}}$, 
we have  straight lines parallel to the real axis  
passing through  the point $w$. In Fig.~3.b we have
plotted these orbits in the complex plane $w$.

\section{A simple example and concluding remarks}

It seems pertinent to conclude by showing the
power of this geometrical approach. In consequence,
we shall analyze in some detail a practical example:
a single glass plate of refractive index $n_1 = 1.5$ 
and thickness $d_1 = 1.1\ $ mm embedded in air.
The plate is illuminated with a monochromatic
light of wavelength \textit{in vacuo} $\lambda = 546 \ $~nm 
that impinges from both ambient and substrate
at an angle $\theta_0 = 45^\circ$.

For this system a standard calculation gives
the following reflection and transmission
coefficients:
\begin{eqnarray}
\label{RT}
R_{as} & = & 
\frac{r_{01} [1 - \exp(- i 2 \beta_1) ]}
{1- r_{01}^2  \exp(- i 2 \beta_1) } , \nonumber \\
& & \\
T_{as} & = & 
\frac{(1 - r_{01}^2) \exp(- i \beta_1)}
{1- r_{01}^2  \exp(- i 2 \beta_1) } , 
\nonumber 
\end{eqnarray}
where $r_{01}$ is the Fresnel reflection coefficient
at the interface 01 (which applies to both $p$ and
$s$ polarizations by the simple attachment of
a subscript $p$ or $s$) and $\beta_1$ is the plate
phase thickness
\begin{equation}
\beta_1 = \frac{2 \pi}{\lambda} n_1 d_1 \cos \theta_1 .
\end{equation}
If we take as initial condition that in the substrate
$z_s = 0.5 \exp(i \pi/6)$, then we obtain from 
Eqs.~(\ref{RT}) the value $z_a = -0.6149 + 0.0498 \ i $
(for $s$ polarization). In Fig.~4 we have plotted 
these points $z_s$ and $z_a$ in the unit disc. 
Obviously, from these (experimental) data alone 
we cannot infer at all the possible path for this 
discrete transformation.

However,  the Iwasawa decomposition remedies 
this serious drawback: from the geometrical
meaning discussed before, and once we know the
values of $\phi$, $\xi$, and $\nu$ [that are easily
computed from Eqs.~(\ref{param})] we get the
intermediate values of $z^\prime$ for the ordered 
application of the matrices $\mathsf{K}(\phi)$,
$\mathsf{A}(\xi)$, and $\mathsf{N}(\nu)$, which,
in fact, determines that the trajectory from $z_s$ to 
$z_a$ is well defined through the corresponding
orbits, as shown in Fig.~4.

Moreover, and this is the important moral we wish
to extract from this simple example, if in some 
experiment the values of $z_s$ and $z_a$ are
measured, one can find, no matter how complicated 
the multilayer is, in a unique way, the three arcs of orbits 
that connect the initial and final points in the unit disc.

We stress that the benefit of this approach lies 
not in any inherent advantage in terms of efficiency
in solving problems in layered structures. Rather, we 
expect that the formalism presented here could provide 
a general and unifying tool to analyze multilayer 
performance in an elegant and concise way that, 
additionally, is closely related to other fields of 
physics, such as special relativity and geometrical optics.

\begin{acknowledgments}
We wish to thank J. Zoido and C. Criado for their help
in computing some of the figures of this paper.
\end{acknowledgments}

\newpage

\begin{figure}
\caption{Wave vectors of the input $[E_{a}^{(+)}$ and  
$E_{s}^{(-)}]$ and output $[E_{a}^{(-)}$ and  
$E_{s}^{(+)}]$ fields in a multilayer sandwiched 
between two identical semi-infinite ambient  and substrate 
media.}
\end{figure}

\begin{figure}
\caption{Unit hyperboloids defined in Eq.~(\ref{intervalo}), 
which represent the space of field states for SO(2,1). 
In each one of them we have plotted a typical orbit
for the matrices: a) $\Lambda_{\mathsf{K}}$,  
b) $\Lambda_{\mathsf{A}}$, and 
c) $\Lambda_{\mathsf{N}}$. In all the figures
we have performed stereographic projection from
the south pole $S$ of the hyperboloid, to obtain the unit
disc in the plane $e^0 =0$ and the corresponding 
orbits, which represent the actions of the 
SU(1,1) matrices: a) $\mathsf{K}$,  b) $\mathsf{A}$, 
and c) $\mathsf{N}$.}
\end{figure}

\begin{figure}
\caption{a) Plot of several orbits in the unit disc 
of the elements of the Iwasawa decomposition 
$\mathsf{K}$,  $\mathsf{A}$,  and $\mathsf{N}$
for SU(1,1) (from left to right, respectively).  
b) Corresponding orbits in the upper complex 
semiplane for the Iwasawa decomposition 
$\mathcal{K}$, $\mathcal{A}$,  and $\mathcal{N}$ 
for  SL(2,$\mathbb{R}$).}
\end{figure}

\begin{figure}
\caption{Geometrical representation in the unit disc
of the action of a single glass plate with the 
parameters indicated in the text. The point $z_s$ 
is transformed by the plate into the point $z_a$.
We indicate the three orbits given by the Iwasawa
decomposition and, as a thick line, the trajectory
associated to the plate action.}
\end{figure}


\begin{thebibliography}{99}
\bibitem{Snell}
When ambient ($0$) and substrate ($m+1$) media are
different, the angles  $\theta _{0}$ and 
$\theta _{m+1}$ are conected by Snell law $n_0\sin \theta_0= 
n_{m+1}\sin \theta_{m+1}$, where 
$n_{j}$ denotes the refractive index of the $j$--th medium.
\end{thebibliography}
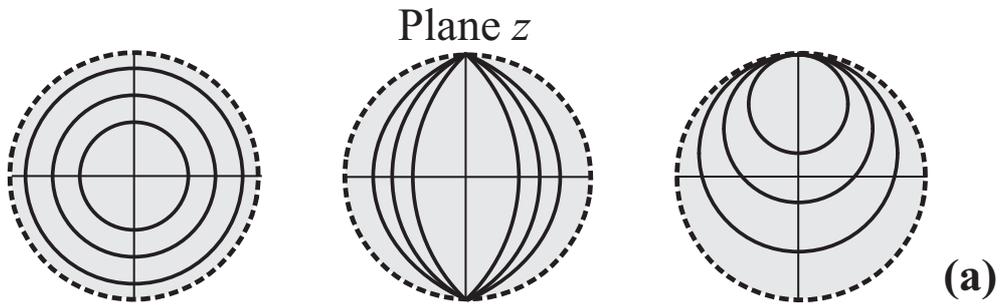
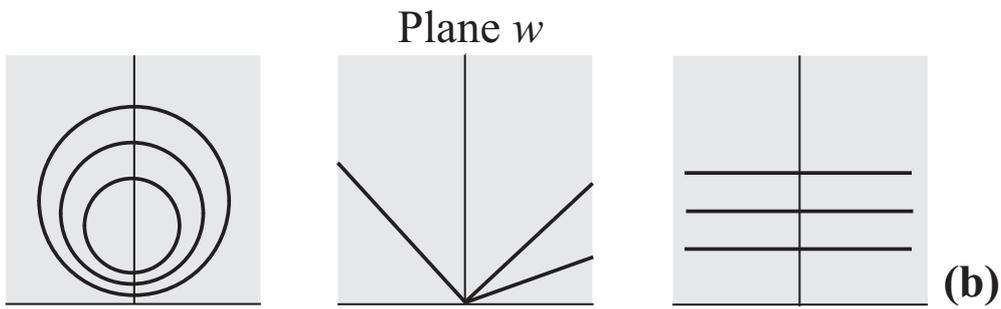
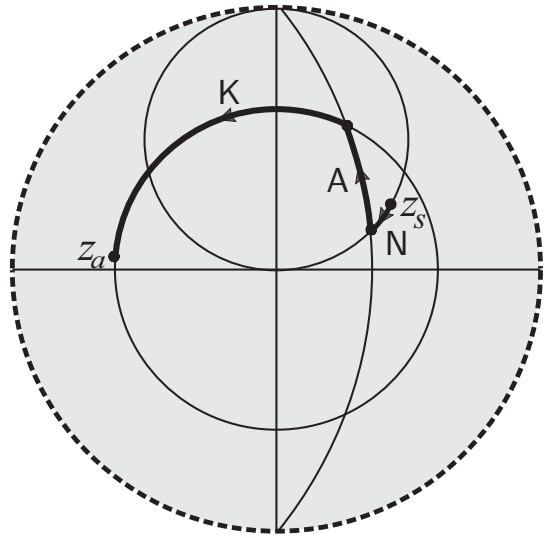
\end{document}